\documentclass{elsart5p}

\usepackage{graphicx}

\usepackage{glas}
\begin{document}
\begin{titlepage}{GLAS-PPE/2008-04}{10$^{\underline{\rm{th}}}$ April 2008}

\title{Alignment of the LHCb Vertex Locator}

\author{M. Gersabeck$^1$\\
        on behalf of the LHCb VELO group\\
\\
$^1$ University of Glasgow, Glasgow, G12 8QQ, Scotland}

\begin{abstract}
LHCb will commence data taking as the first dedicated heavy flavour experiment at a hadron collider in 2008.
A very high hit precision from its vertex detector (VELO) is essential to meet the tight requirements of vertex reconstruction in B-physics.
The single hit precision  of the VELO is better than $10~\mathrm{\mu{}m}$.
However, the VELO is operated only $8~\mathrm{mm}$ from the beam and must be retracted and reinserted each LHC fill.
Hence, the detector places unique demands on its alignment algorithm.
The partially assembled VELO system has already been tested in a beam test.
The novel software alignment methods are presented together with their interplay with the metrology measurements.
Results from Monte Carlo simulation studies are discussed and recent beam test results are shown that prove the method's precision at the micron level.

\vspace{0.5cm}
\begin{center}
{\em $10^{th}$ International Conference On Instrumentation For Colliding Beam Physics}\\
{\em Novosibirsk, Russia}
\end{center}
\end{abstract}

\newpage
\end{titlepage}

\begin{frontmatter}


 \title{Alignment of the LHCb Vertex Locator}
 \author{M.~Gersabeck}
 \corauth[cor1]{}
 \address{Department of Physics and Astronomy, University of Glasgow, Glasgow, G12 8QQ, UK\\
          on behalf of the LHCb VELO group}





\begin{abstract}
LHCb will commence data taking as the first dedicated heavy flavour experiment at a hadron collider in 2008.
A very high hit precision from its vertex detector (VELO) is essential to meet the tight requirements of vertex reconstruction in B-physics.
The single hit precision  of the VELO is better than $10~\mathrm{\mu{}m}$.
However, the VELO is operated only $8~\mathrm{mm}$ from the beam and must be retracted and reinserted each LHC fill.
Hence, the detector places unique demands on its alignment algorithm.
The partially assembled VELO system has already been tested in a beam test.
The novel software alignment methods are presented together with their interplay with the metrology measurements.
Results from Monte Carlo simulation studies are discussed and recent beam test results are shown that prove the method's precision at the micron level.
\end{abstract}

\begin{keyword}
Alignment \sep LHCb \sep VELO \sep Millepede

\PACS 
\end{keyword}
\end{frontmatter}

\section{Introduction}
LHCb is due to commence data taking as the first dedicated B-physics experiment at a hadron collider in the second half of 2008.
One of the key characteristics of measurements of B-meson decays is their displaced decay vertex.
This requires particularly precise measurements of both primary and secondary vertex positions.
For LHCb, this is performed by the Vertex Locator (VELO)~\cite{bib:Olaf-08}.

With a single hit resolution of better than $10~\mathrm{\mu{}m}$~\cite{bib:ReOpt-03}, even for tracks with minimal charge sharing, a knowledge of the VELO sensor positions on the micron level is vital.
Studies have shown that for example a misalignment of the VELO sensors of the order of the single hit resolution deteriorates the precision on the B-meson proper time by about $50\%$~\cite{bib:Marco-08}.

A challenge to the determination of alignment constants is posed by the movement of the two halves away from the beam during injection.
This means that more than once a day each VELO half is moved by $3~\mathrm{cm}$ each way.
Each time the vertex locator is moved a new alignment is required.

In section 2 the techniques for alignment are explained in detail.
In the third section the performance of the alignment algorithms determined from a beam test is discussed and followed by the conclusion in section four.

\section{VELO alignment method}
The LHCb VELO consists of two retractable halves with each half having $21$ modules that each support two semi-circular silicon strip sensors, one measuring $R$ (radial) and the other $\Phi$ (azimuthal) co-ordinates~\cite{bib:ReOpt-03}.
When in their nominal position, the sensors form a series of disks, of radius $42$~mm, placed normally to the beam axis.
A cut out up to a radius of $7$~mm allows for the beams to pass through the detector without directly interacting with the sensors.
However, the LHC filling procedure requires that the VELO is retracted by at least $30$~mm from its nominal operating position.
Thus, during injection the two detector halves are separated by $\approx{}60$~mm.

After a precise assembly and a detailed metrology, the position of the active parts of the VELO during operation will be determined by track based software alignment algorithms~\cite{bib:Seb-08}.
This procedure is divided into three stages.
First, the relative alignment of the two sensors on each module are determined.
This is followed by the alignment of all modules in each of the two halves with respect to each other.
Finally, the relative position of the two halves is measured.

All algorithms exploit measurements of track residuals.
Here, residuals are the difference between the intercept of a fitted track with a sensor and the measured hit position on this sensor.
A systematic bias of the residual distribution for one sensor can then be associated with a misalignment.

For the relative alignment of the two sensors on each module it is important to note that residuals are only sensitive to misalignments perpendicular to the direction of the strip.
As the strip direction varies across the sensor, this produces a relationship between the residual distribution across the sensor to its displacement as follows:
\begin{eqnarray}
  residual_{R} & = & -\Delta{}x \cos\phi_{track} + \Delta{}y \sin\phi_{track}\\
  residual_{\Phi} & = & \Delta{}x \sin\phi_{track} + \Delta{}y \cos\phi_{track} + \Delta\gamma r_{track},
\end{eqnarray}
with $\Delta{}x$ and $\Delta{}y$ the translation misalignment along the $x$ and $y$ axes respectively, and $\Delta\gamma$ the rotation misalignment around the $z$ axis.
For each sensor the residual distribution is determined from a sample of about 20000 tracks and then fitted according to these equations.
The difference in the fit result gives the relative misalignment of the two sensors.

Both the relative module alignment and the VELO half alignment use a global minimisation method based on the {\tt Millepede} algorithm~\cite{bib:Blobel-07}.
This procedure relies on a set of linear equations relating a track model to residuals, where the latter are expressed as a function of alignment constants ($\Delta$) and their derivatives ($c$).
Summing up over all tracks ($tr$) and dimensions of the track model ($i$) one can derive a global $\chi^{2}$ function:
\begin{equation}
  \chi^{2} = \sum_{tr,i}\left(y^{tr,i} - \sum_{j} x^{tr,i}_{j} \cdot \alpha^{tr}_{j} - \sum_{k} c^{tr,i}_{k} \cdot \Delta_{k}\right)^{2}/\sigma_{i}^{2},
\end{equation}
with the measurements ($y$), the track parameters ($\alpha$) and their derivatives ($x$).
The minimised $\chi^{2}$ can then be written as a matrix equation.
\begin{equation}
  V \cdot \left(~^{\Delta}_{\alpha_{track}}\right) = \left(~^{f(C,Y,\sigma)}_{f(X,Y,\sigma)}\right)
\end{equation}
Inverting the matrix $V$ directly yields the optimal set of alignment constants ($\Delta$).

However, this matrix is of the size of the number of global parameters\footnote{for the modules in one VELO half: $n_{global}=n_{module}\times{}n_{d.o.f.}=21\times{}6$} plus the number of tracks times the number of parameters per track.
As one needs of the order of many tens of thousands tracks to align the modules in a VELO half, inverting a matrix of this size is non-trivial.
The {\tt Millepede} algorithm exploits the fact that this matrix is very sparse as its major block has only small blocks for each track along the diagonal.
Using this knowledge, the whole procedure is able to align the modules inside each of the VELO halves within a few minutes on a single CPU\footnote{1 CPU = 1000 SpecInt2000 units}.
Furthermore, it intrinsically takes into account the correlations between the different modules and includes the functionality for constraining weak degrees of freedom, i.e. deformations that are difficult or impossible to unfold with tracks. 

To align the two VELO halves with respect to each other, the same algorithm is used.
However, the track sample used here consists of tracks that produce hits in both VELO halves.
This is possible due to a small region in the $x$-$y$ view (around 2\% of the active area) where sensors from both sides of the VELO overlap.
As these tracks produce a very stringent constraint on the relative position of the two halves, only a few hundred are needed to determine the alignment constants.
These are expected to exist in sufficient quantities of either muon beam halo tracks or tracks from beam-gas interactions.
As an alternative, tracks from a common primary vertex can be used to align the two halves.
However, due to the additional degrees of freedom of the vertex fit, these only provide a weaker handle on the alignment parameters.

During data taking it is foreseen to run both the module alignment and the VELO half alignment algorithms after each fill when the two halves have reached their closed position.
This process will take about $10$ to $15$ minutes and will thus ensure a high precision of the alignment already for the first processing of the data which starts with a delay of a few hours with respect to the data taking.
The relative sensor alignment is not expected to change during the VELO movements and will hence be performed at larger time intervals.
However, all alignment constants will be continuously monitored.

\section{Beam test results}

A first test of the alignment methods using real data was performed after a beam test in late 2006.
The beam test setup comprised ten VELO modules and used the same hardware and software readout chain like the final experiment.
During different phases of this test different sets of six out of the ten VELO modules were read out simultaneously.

The beam used was a $180$~GeV secondary pion beam which was directed either directly onto the sensors or aimed at small lead targets that produced secondary tracks to be detected by the VELO sensors.
The targets were located at various positions along the position of the nominal beam line.

\begin{figure}
  \includegraphics[width=\columnwidth]{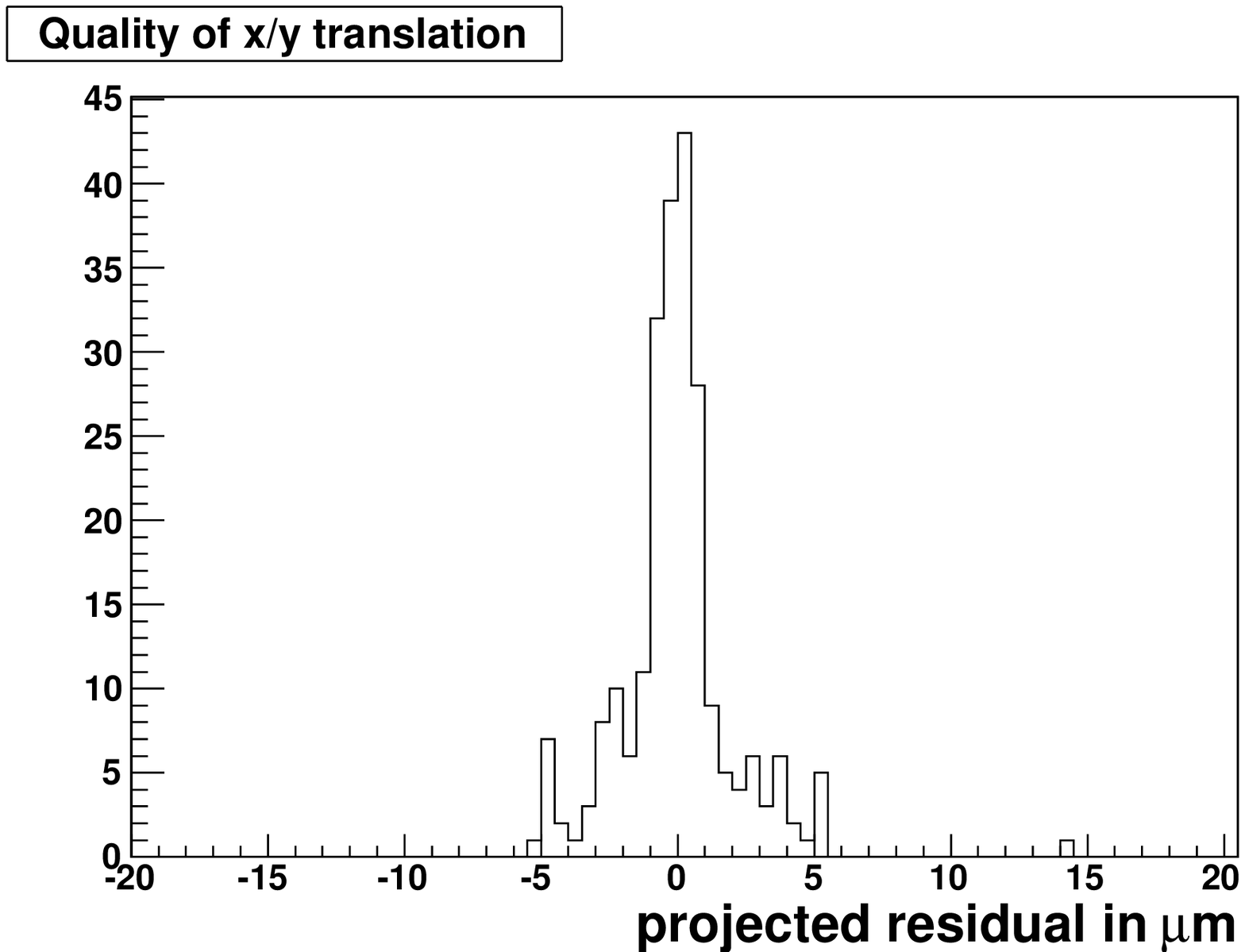}
  \caption{Alignment quality plot showing the projection of the residual distribution vs $\phi$. The width of the distribution of $0.9~\mathrm{\mu{}m}$ shows the size of the remaining translational misalignments.}
  \label{fig:aliQxy}
\end{figure}

A comprehensive overview of the performance of the alignment software with beam test data is given in Ref.~\cite{bib:Marco-08b}.
A first indication of the quality of the alignment algorithms can be obtained by studying residual distributions.
As described in the previous section, the amplitude of any observed sinusoidal distribution of the residuals as a function of the azimuthal angle $\phi$ is directly related to the size of the translational misalignment of the sensors.
To get a more quantitative measure, the shape of these residual distributions has been projected to the residual axis for all sensors, yielding a one-dimensional distribution whose width is related to the size of the remaining misalignments.
Figure~\ref{fig:aliQxy} shows this distribution and the width of only $0.9~\mathrm{\mu{}m}$ confirms that an alignment precision on the micron level has been achieved.

\begin{figure}
  \includegraphics[width=\columnwidth]{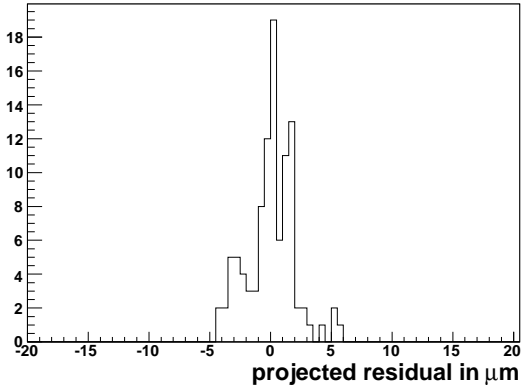}
  \caption{Alignment quality plot showing the projection of the residual distribution vs $r$. The width of the distribution of $2.0~\mathrm{\mu{}m}$ shows the size of the remaining misalignments, translating into a remaining rotational misalignment of $0.1$~mrad.}
  \label{fig:aliQg}
\end{figure}

In the same way, the slope of the residual distribution as a function of the radial co-ordinate $r$ gives the size of the misalignment due to a rotation around the $z$ axis.
Again, its projection to the residual axis can be related to the overall size of the remaining misalignments.
Figure~\ref{fig:aliQg} shows the distribution of the projected residual distributions vs $r$ for all $\Phi$ sensors.
The width of $2.0~\mathrm{\mu{}m}$ can be related via an effective radius to the precision on rotational misalignments, yielding a size of remaining misalignments of $0.1$~mrad.

Compared to the single hit resolution for the smallest strip pitch and for tracks with minimal charge sharing of roughly $9~\mathrm{\mu{}m}$ these levels of alignment precision lead to the conclusion that no deterioration in the detector's performance has to be expected due to misalignments.

One effect that can potentially limit the accuracy of the alignment algorithms and thus the single hit precision is asymmetric cross-talk introducing a bias depending on the readout order of the single strips.
Particularly in the case of the $R$ measuring sensors, this effect is strongly visible and can lead to a bias of several microns.
It originates from the fact that the readout direction is such that, for some parts of the sensor, strips are read out towards increasing radii while, in other parts of the sensor, strips with smaller radius are read out later.
A forward cross-talk introduced in long-distance cables, connecting the detector with the readout electronics, hence leads to a bias either towards a smaller or larger radius.

To compensate for this, a finite impulse response filter has been put into place.
This algorithm, when performed during the digitisation of the raw data, has been shown to be able to correct for the cross-talk to a level that does no longer limit the alignment algorithms.

\begin{figure}
  \includegraphics[width=\columnwidth]{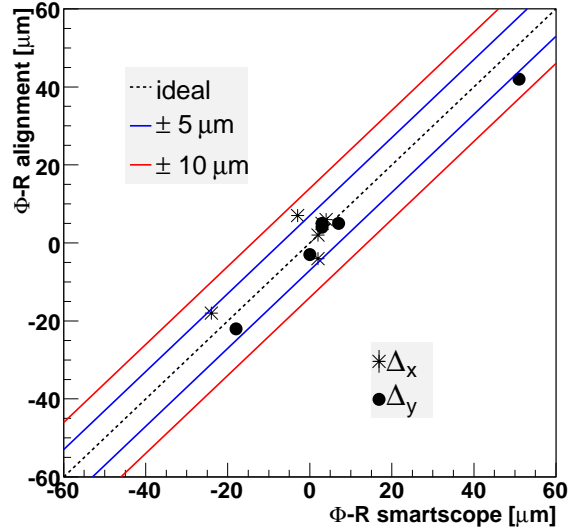}
  \caption{Comparison of alignment constants determined by software with thus determined by a smartscope metrology.}
  \label{fig:ali_sm}
\end{figure}

A completely independent test of the quality of the alignment software has been performed using beam test data by comparing the alignment constants as obtained by the software algorithms to those measured by the module metrology during assembly~\cite{bib:Tor-07}.
Figure~\ref{fig:ali_sm} shows this comparison for the relative translation of $R$ and $\Phi$ sensors on six modules.
It shows that both methods are in agreement down to a level of $5~\mathrm{\mu{}m}$ which is comparable to the expected accuracy of the smartscope metrology, thus confirming the precision of the software alignment.

\begin{figure}
  \includegraphics[width=\columnwidth]{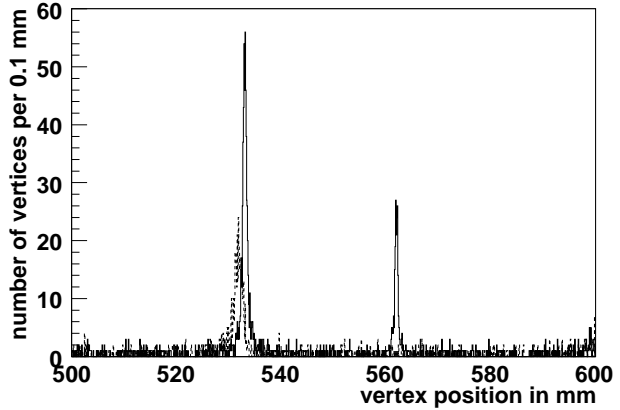}
  \caption{Reconstructed vertices as function of their position along the beam axis without (dashed) and with correction of misalignments.}
  \label{fig:vertices}
\end{figure}

An obvious example for the need of alignment algorithms turned out to be the reconstruction of vertices of tracks from interactions of the beam in the targets.
Despite having two targets in a certain region traversed by the beam, only vertices from one target were reconstructed when assuming the nominal geometry of the sensors.
After correcting for misalignments with the output of the alignment algorithms, not only vertices from the second target were reconstructed but also the vertex resolution showed a clear improvement (see Fig.~\ref{fig:vertices}).
In general, precise vertexing is vital for LHCb due to the tiny displacement of the $B$ decay vertices that have to be reconstructed already at trigger level.

Finally, another valuable cross-check was the study of the stability of the VELO alignment.
Therefore, alignment constants have been determined while the detector was operated in air at atmospheric pressure as well as during operation in a vacuum of $10^{-3}$~mbar.
As no significant change in the alignment constants has been observed it was confirmed that there is no sizable deformation due to effects like out-gassing during evacuation of the detector volume.

Another two sets of alignment constants that have been compared were determined from data taken before and after the detector had been moved, similar to the movement during each fill of the LHC.
Here as well, no significant change in the alignment constants has been seen.
It may be concluded that the detector alignment, and hence its performance, is not compromised by the need to move the detector between fills.

\section{Conclusion}
The VELO software alignment has been presented with its three steps: the relative alignment of the sensors, using the shape of residual distributions, and the module and VELO half alignment using the global $\chi^{2}$ minimisation algorithm {\tt Millepede}.
With these algorithms all necessary alignment constants can be determined within minutes on a single CPU.

It has been shown that with a precision at the micron level, the algorithm uncertainties are well below the single hit resolutions.
Hence, the alignment will not introduce adverse systematic effects on physics observables.
An excellent agreement of software results with metrology data underlines the precision in the determination of the alignment constants.
Finally, a good stability of the VELO modules under movements and evacuation of the detector volume has been demonstrated.



\end{document}